\documentclass[prl,aps,twocolumn,showpacs,floatfix]{revtex4}
\usepackage{graphicx}
\usepackage{amsmath}
\usepackage{amssymb}
\usepackage{bm}

\newcommand{\hP}{\hat{P}}
\newcommand{\hth}{\hat{\theta}}
\newcommand{\hH}{\hat{H}}
\newcommand{\bGam}{{\bm \Gamma}}
\newcommand{\bk}{{\bm k}}
\newcommand{\ba}{{\bm a}}
\newcommand{\br}{{\bm r}}

\newcommand{\bd}{{\bm d}}
\newcommand{\hT}{\hat{T}}
\newcommand{\bkappa}{{\bm \kappa}}
\newcommand{\hsig}{\hat{\sigma}}
\newcommand{\hSig}{\hat{\Sigma}}

\begin{document}

\title{$\mathbb{Z}_2$ Topological Insulators in Ultracold Atomic Gases}

\author{B. B{\' e}ri and N. R. Cooper}
\affiliation{T.C.M. Group, Cavendish Laboratory, J.~J.~Thomson Ave., Cambridge
  CB3~0HE, United Kingdom.}

\begin{abstract}

  We describe how optical dressing can be used to generate bandstructures for
  ultracold atoms with non-trivial $\mathbb{Z}_2$ topological order.  Time
  reversal symmetry is preserved by simple conditions on the optical
  fields. We first show how to construct optical lattices that give rise to
  $\mathbb{Z}_2$ topological insulators in two dimensions. We then describe a
  general method for the construction of three-dimensional 
  $\mathbb{Z}_2$ topological insulators. 
A central feature of our approach is a new way to understand  $\mathbb{Z}_2$ topological insulators starting from
the nearly-free electron limit.

\end{abstract}
\date{May 6, 2011}

\maketitle

Recently it has been realized that the electronic structure of
materials is much richer than had previously been thought.  The
apparently simple band insulator
admits an intricate topological classification, and can behave as a
``topological insulator'' which is gapped in the bulk but carries
gapless, metallic, states on its surface\cite{hasankane,qizhang}.
The classic example is the integer quantum Hall effect (IQHE) of a
two-dimensional (2D) electron system in a magnetic field. In this
setting, of 2D with broken time-reversal symmetry (TRS), band
insulators are characterized by the integer-valued Chern
number\cite{thoulesschern}, which determines the number of metallic
surface states. Presently, interest is focused on settings that
preserve TRS, since unlike the IQHE, topological insulators are possible in both 2D and 3D. They are classified by a $\mathbb{Z}_2$ invariant\cite{KaneMele05FuKaneMele07,MooreBalents,Roy2dRoy3d}: band insulators 
either are trivial, or are non-trivial and support a metallic surface
state with interesting spin-structure.

While the effects of interactions are rather well understood in the IQHE case, our understanding of strong correlations in $\mathbb{Z}_2$ topological insulators, especially in 3D is yet in its infancy. 
Ultracold atomic gases might hold much promise in this direction:
they allow the effects of interactions and other perturbations to be explored in a controlled way\cite{blochdz,lewensteinrev}.  To take advantage of this in the topological insulator context, it is a crucial prerequisite to find a way to simulate bandstructures with nontrivial $\mathbb{Z}_2$ invariant. The purpose of this Letter is to solve this problem in both 2D and 3D in a simple, generic  and experimentally feasible fashion.  

The standard technique for imposing bandstructures on ultracold atom
gases is the optical lattice: a periodic scalar potential formed from
standing waves of light\cite{blochdz}.  Optical lattices lead to simple bands with no topological character.  In recent work\cite{ofl}, it has been shown
that by using  the optical fields to couple more than one internal atomic
level, leading to effective gauge fields\cite{dalibardreview},
 one can generate ``optical flux lattices'' which break TRS and
lead to bands  with non-zero Chern number. Here we show that, by suitable optical dressing, one can form optical lattices that 
generate non-abelian gauge fields\cite{lewensteinrev,dalibardreview}, in a way that
preserves TRS and leads to bands with
non-trivial $\mathbb{Z}_2$ invariant. Our approach is related to methods using optical fields to simulate 
the effects of spin-orbit coupling
\cite{dalibardreview,dudarevstanescuspielmanso}.  It is however very different to  existing proposed cold-atom
implementations of $\mathbb{Z}_2$ topological
insulators\cite{goldmantopoins}, which imprint non-abelian gauge
potentials on atoms in a deep optical lattice. 
Our key innovation is to approach the problem from the opposite, nearly free electron limit. This allows for large gaps, with size set by the recoil energy. 
 Moreover, our proposal can be implemented with a relatively simple laser set-up.
In addition to its advantages to cold atom realizations, our nearly free electron approach brings a qualitatively new viewpoint to the field of topological insulators.

We consider an atom with position ${\bm r}$ and momentum ${\bm p}$ and
with $N$ internal states which is described by the Hamiltonian
\begin{equation} \hat{H} = \frac{\hat{{\bm
      p}}^2}{2m} \openone_N + V \hat{M}({\bm r}), \label{eq:ham}
\end{equation}
where $V$ has dimensions of energy. $\openone_N$ is the identity,
while $\hat{M}({\bm r})$ is a position-dependent $N\times N$ matrix
acting on the internal states of the atom.  To realize a
$\mathbb{Z}_2$ topological insulator requires $N$ to be even, owing to
a ``spin'' structure, and the Hamiltonian has to be invariant under
time reversal, $\hat{\theta}=i\hat{\sigma}_y\hat{K}$\cite{hasankane,qizhang}. (We denote by
$\hsig_i$ the Pauli matrices in spin space, and by $\hat{K}$ complex
conjugation.)  This requires that $ \hat{M} = \hat{\theta}^{-1}
\hat{M}\hat{\theta}$.  The smallest non-trivial case has $N=4$ with
\begin{equation}
 \hat{M} = 
\left(\begin{array}{cc}
(A+B)\openone_2 & C\openone_2 - i\hat{\vec{\sigma}}\cdot\vec{D}\\
C\openone_2 + i\hat{\vec{\sigma}}\cdot\vec{D} &   (A-B)\openone_2
\end{array}\right).
\label{eq:m}
\end{equation}
Here $A,B,C$ are  real numbers, and $\vec{D} = (D_x,D_y,D_z)$  a real
three-component vector. $\hat{M}$ can be also written as 
\begin{equation}
\hat{M}=A\openone_4+B\hSig_3+C\hSig_1+\vec{D}\hSig_2\hat{\vec{\sigma}},
\end{equation}
where $\hSig_j$ are the Pauli matrices in the block structure of Eq.~\eqref{eq:m}.

We propose to realize Hamiltonians of this form using four internal
states of an atom, with the entries in Eq.~\eqref{eq:m} provided by
optical fields\cite{dalibardreview}.  Many implementations are
possible.  In principle the various transitions between the internal
states could be spectrally resolved, allowing the freedom separately
to introduce coupling lasers for all transitions (or pairs of
lasers for Raman coupling).
Our results are very general, and could be applied to any such
implementation.  However, to make the discussion concrete, we shall
focus on a particular realization in which the coupling laser operates
on a single frequency, and therefore has very limited freedom.  

The implementation we consider makes use of the properties of the
groundstate ($^1S_0 = g$) and long-lived excited state ($^3P_0=e$) of ytterbium (Yb) or
an alkaline earth atom.
The usefulness of these two levels for optically induced gauge fields
was pointed out by Gerbier and Dalibard\cite{gerbier}: the long
lifetime of the excited state reduces spontaneous emission;
furthermore, there
is a convenient
``anti-magic'' wavelength $\lambda_{\rm am}$ at which the two states
experience scalar optical potentials of opposite signs,  allowing
state-dependent potentials, $\pm V_{\rm am}({\bm r})$, to be
implemented.
We focus on $^{171}$Yb\cite{taie} which has nuclear spin $I=1/2$. Then
both $g$ and $e$ have two internal states, and there are four states
in total; these will correspond to the $4\times 4$ structure in Eq.~\eqref{eq:m}.
 We consider the magnetic field to be sufficiently small that the Zeeman
splitting is negligible, and all four $e$-$g$ transitions involve the
same frequency $\omega_0 = (E_e-E_g)/ \hbar$. 
Combining the single-photon coupling, with electric
field $\vec{E} = \vec{\cal E} e^{-i\omega t} + \vec{\cal E}^*
e^{i\omega t}$, with the state dependent potential $V_{\rm am}$, leads
to the optical potential which in the rotating wave
approxmation\cite{cohen} is
\begin{equation}
 V \hat{M}
= \left(\begin{array}{cc}
\left(\frac{\hbar}{2}\Delta +V_{\rm am}\right)\openone_2 & -i\hat{\vec{\sigma}}\cdot\vec{{\cal E}}d_r\\
 i\hat{\vec{\sigma}}\cdot\vec{{\cal E}}^*d_r & -\left(\frac{\hbar}{2}\Delta +V_{\rm am}\right)\openone_2
\end{array}\right),
\label{eq:coupling}
\end{equation}
where $\Delta = \omega -\omega_0$ is the detuning.  The form of the
off-diagonal term follows from the Wigner-Eckart theorem, with $d_r$
the reduced dipole moment\cite{llqm}. 
A comparison of Eq.~\eqref{eq:coupling} with Eq.~\eqref{eq:m} shows that
the optical coupling (\ref{eq:coupling}) describes a TRS situation
provided all components of $\vec{{\cal E}}$ have the same phase, in
which case $\vec{{\cal E}}$ can be chosen real.

{\it Two dimensional systems.} We consider first cases in which the atoms are
tightly confined in the $z$-direction, so their motion in the $x-y$ plane is (quasi)-two-dimensional.  A $\mathbb{Z}_2$ topological insulator can
be formed by choosing the electric field, detuning and state-dependent
potential in Eq.~\eqref{eq:coupling} such that
\begin{eqnarray}
\label{eq:efield}  d_r \vec{\cal E}   & = &  V \left(\delta,  \cos({\bm r}\cdot{\bm \kappa}_1), 
    \cos({\bm r}\cdot{\bm \kappa}_2) \right)\\
\label{eq:stdep}
  \frac{\hbar}{2}\Delta + V_{\rm am}({\bm r})  & = & V \cos[{\bm r}\cdot ({\bm \kappa}_1+{\bm \kappa}_2)]
\end{eqnarray}
with ${\bm \kappa}_1 = (1,0,0)\kappa$, ${\bm \kappa}_2 =
(\cos\theta,\sin\theta,0)\kappa$.  The amplitudes are chosen to have a
common energy scale $V$, which can be interpreted as a measure of the
Rabi coupling. This energy scale should be compared to the
characteristic kinetic energy scale, the recoil energy $E_R \equiv
\hbar^2\kappa^2/2m$.  The interspecies coupling (\ref{eq:efield}) is
formed from three (standing) waves of linearly polarised light at the
coupling frequency $\omega$: two of equal amplitude with wavevectors
in the 2D plane (${\bm \kappa}_1$ for $y$-polarization and ${\bm
  \kappa}_2$ for $z$-polarization) and one with wavevector normal to
the 2D plane for $x$-polarization with an amplitude smaller by a
factor $\delta$. Since $\omega \simeq \omega_0$ the magnitude of the
in-plane wavevectors is $\kappa \simeq 2\pi/\lambda_0$ with
\mbox{$\lambda_0=578$nm} the wavelength of the $e$-$g$ transition. The spatial
dependence of $V_{\rm am}$ is set by a standing wave at the anti-magic
wavelength $\lambda_{\rm am}$\cite{gerbier}, which creates a
state-dependent potential with $|\bkappa_1+\bkappa_2| =
4\pi/\lambda_{\rm am}$.  This fixes the angle $\theta =
2\arccos\left(\pm {\lambda_0/\lambda_{\rm am}}\right)$. For Yb,
$\lambda_0/\lambda_{\rm am}\simeq 1/2$ (to an accuracy of about
3$\%$), so $\theta \simeq \pm2\pi/3$. For simplicity, in all following
discussions we fix $\theta = 2\pi/3$ and define $a \equiv
4\pi/(\sqrt{3}\kappa)$.  The optical coupling $\hat{M}$ then has the
symmetry of a triangular lattice with lattice vectors ${\bm a}_1 =
(\sqrt{3}/2,-1/2)a$ and ${\bm a}_2 = (0,1)a$.

The physics arising from this form of optical coupling is most clearly
exposed by applying a unitary transformation $\hat{U}=2^{-1/2}(\openone_4-i\hSig_3\hsig_2)$\cite{Utrnote} to the dimensionless
coupling $\hat{M}$
\begin{eqnarray}
\hat{M}^\prime =\hat{U}^\dag \hat{M} \hat{U} & = & 
c_1\hSig_1 +c_2\hSig_2\hsig_3+c_{12}\hSig_3+\delta\hSig_2\hsig_1.
\label{eq:umu}
\end{eqnarray}
We have used the shorthand $c_i \equiv \cos{\bm r}\cdot {\bm \kappa}_i$
and $c_{12}\equiv \cos[{\bm r}\cdot ({\bm \kappa}_1+{\bm \kappa}_2)]$. 
For $\delta =0$ this matrix decouples into two $2\times2$ blocks, one
for each eigenvalue of $\hsig_3$, as does the Hamiltonian since the
kinetic energy (\ref{eq:ham}) is diagonal.  Thus, the four-level
system decouples into two two-level subsystems, each
experiencing its own form of optical dressing, $c_1\hSig_1 \pm
c_2\hSig_2+c_{12}\hSig_3$, differing only in the sign of the $c_2$
term.  This is precisely the optical dressing 
required to realize the ``triangular'' optical flux lattice described
in Ref.\cite{ofl} (See \cite{kplusfootnote}).
This optical flux lattice causes the atoms to experience a (periodic)
effective magnetic field with $N_\phi=2$ flux quanta in the unit cell.
For $V \gtrsim 0.2 E_R$ the lowest energy band (which
is twofold degenerate\cite{ofl}) is separated in energy from
higher bands. This band is topologically non-trivial, having a Chern
number of $\pm 1$, with sign set by the net sign of the coefficients
$c_1, c_2, c_{12}$.  Since this sign differs for the two subsystems,
their lowest bands have equal and opposite Chern numbers. (This is
required by the fact that these subsystems are related by
time-reversal.)  Thus, for $\delta=0$ the optical dressing realizes
the bandstructure required to generate the quantum spin Hall (QSH)
effect: there are two decoupled subsystems, the lowest energy bands of
which have Chern numbers $\pm 1$.  Each of these bands is filled when
the 2D density of fermions (in that component) is equal to the
magnetic flux density, $N_\phi/(\sqrt{3}/2 a^2) =
4/(\sqrt{3}a^2)$. When both bands are filled, at a total fermion
density of $n_{\rm 2D} = 8/\sqrt{3}a^2$, the bulk of the system is
gapped.  However, there will be a metallic surface state involving
counterpropagating edge modes of opposite spin. This metallic surface state
is the hallmark of the topological order that characterizes the QSH
effect.

\begin{figure}
\includegraphics[width=\columnwidth]{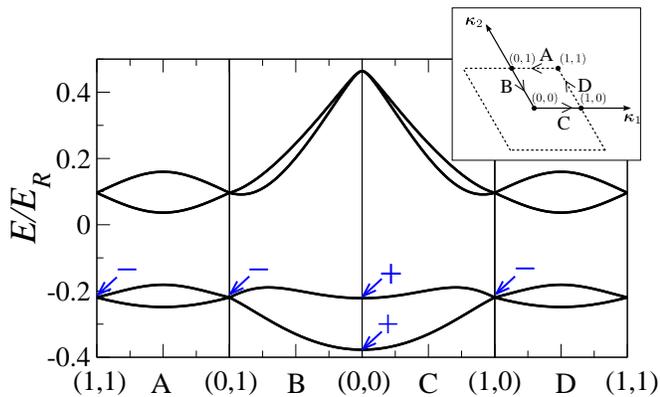}
\caption{ \label{fig:triband} Lowest energy bands for the 2D
  topological insulator with $V = 0.5 E_R$ and $\delta = 0.25$ for wavevectors on the perimeter of the  parallelogram (shown in the inset) with corners $\bGam_{nm}\!=\!(n\bkappa_1+m\bkappa_2)/2$ ($n,m\!=\!0,1$).    Each band is
  twofold degenerate, with fourfold degeneracies at $\bGam_{nm}\neq0$. The four bands with $E/E_R \lesssim -0.1$ have
  non-trivial $\mathbb{Z}_2$ invariant, as follows from the eigenstate parities ($\pm$) at $\bGam_{nm}$.}
\label{fig:split}
\end{figure}

The QSH effect is a special case of the $\mathbb{Z}_2$ topological insulator.
The $\mathbb{Z}_2$ topological insulator, in general, does not allow
separation into two decoupled systems, but retains counterpropagating edge
states that are protected only by TRS. A nonzero value of $\delta$ introduces
a coupling between the two spin-states in a manner that
preserves TRS, in direct analogy to the effects of spin-orbit coupling on the
electronic structure of solids.  In Fig.\ref{fig:split} we show the few lowest
energy bands for a small nonzero value of $\delta$. (The bands were calculated by numerical diagonalization in the plane wave basis.) 
For $\delta =0$ all bands are fourfold degenerate. For $\delta\neq 0$ this
degeneracy splits into two doubly degenerate branches at most wavevectors.
However, the fourfold degeneracy remains at three nonequivalent symmetry points in $k$-space:  $\bkappa_1/2$,  $\bkappa_2/2$, and  $(\bkappa_1+\bkappa_2)/2$. (The reciprocal lattice vectors are simply given by $\bkappa_i$.)

Due to the coupling between the two subsystems, the Chern number of a subsystem is no longer well defined, but the bands still carry a topological character since the $\delta=0$ case is adiabatically connected to the QSH limit.   To demonstrate this, one has to use a more general $\mathbb{Z}_2$ invariant\cite{FuKaneinv}. This invariant takes a particularly simple form for our system owing to its symmetry under spatial inversion,  $\hP:\br\rightarrow -\br$. Inversion symmetric insulators, as shown by Fu and Kane\cite{FuKaneinv,supplmat}, are $\mathbb{Z}_2$ nontrivial if 
\begin{equation} \prod_{n,m=0,1}\prod_{\alpha\in\text{filled}}
  \xi^{(\alpha)}_{nm}=-1,
\label{eq:FuKaneinv}
\end{equation}
where $\xi^{(\alpha)}_{nm}$ are the inversion eigenvalues corresponding to the $\alpha$-th Kramers doublet of bands  at the momenta ${\bm\Gamma}_{nm}=(n\bkappa_1+m\bkappa_2)/2$ ($n,m=0,1$). Evaluating Eq.~\eqref{eq:FuKaneinv} for our system we find that it is indeed in a topological phase: the points with fourfold degeneracies contribute with $(-1)$ while $\bk=0$ contributes with unity, as indicated in Fig~\ref{fig:split}. 
Although we have shown bands for the special case of a triangular lattice, our approach is very general: starting from an optical flux lattice of any symmetry\cite{ofl}, our method will lead to a 2D $\mathbb{Z}_2$ topological insulator.

\begin{figure}
  \includegraphics[width=0.8\columnwidth]{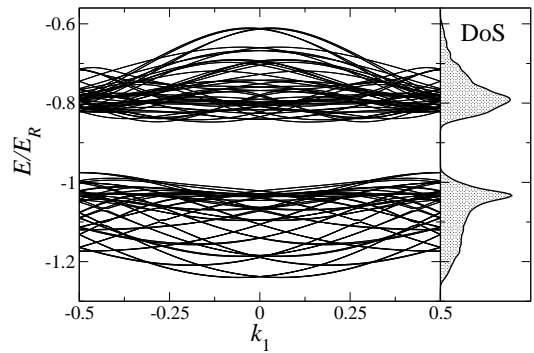}
  \caption{Lowest energy bands for the model
    Eq.~\eqref{eq:cossumpure}, for a set of wavevectors ${\bm k}
    = \sum_i k_i {\bm \kappa}_i$ with $k_2$ and $k_3$ uniformly spaced
    across the Brillouin zone. The parameters are $V=0.9E_R$,
    $\delta=1.0$, $\mu=-0.2$. The four bands with $E/E_R \lesssim
    -0.9$ have non-trivial $\mathbb{Z}_2$ invariant which, together with the  clear band gap, demonstrate the topological insulator phase. The right hand panel shows the density of states.}
\label{fig:3dTI}
\end{figure}

{\it Three dimensional systems.}  
Our approach has the remarkable feature that it generalizes naturally to  3D topological insulators.  One simply needs to substitute
\begin{equation}c_{12}\rightarrow c_{12}+ \delta (\mu +c_{13}+c_{23})\qquad \delta\rightarrow\delta\cos(\bkappa_3\cdot\br),
\label{eq:cossumpure}
\end{equation} 
where our notation follows that of Eq.~\eqref{eq:umu}, and we use $\bkappa_3=(0,0,1)\kappa$. 
That this system is a good candidate for a topological insulator can  be checked    using the parities as in Eq.~\eqref{eq:FuKaneinv}, but the product is now over eight momenta $\bGam_{nml}=(n\bkappa_1+m\bkappa_2+l\bkappa_3)/2$ ($n,m,l=0,1$)\cite{FuKaneinv}. Evaluating the product shows that the parities are those of a $\mathbb{Z}_2$ topological phase for a large range of parameters. Importantly, this range includes systems which are insulators as well. 
This is demonstrated in Fig.~\ref{fig:3dTI}, where we show the few lowest
energy bands of the model for a point in the regime where the parity product is negative. A clear
bandgap separates the lowest four bands from the higher bands, showing that the system is an insulator. 
This extension to 3D entails additional geometrical constraints on the
wavevectors of the coupling and state-dependent lasers. For example,
for ${\bm \kappa}_i \perp {\bm \kappa}_j $  the
state-dependent potential $c_{ij}$ has to have periodicity
$\lambda_0/\sqrt{2}$.  For Yb, this differs from that achieved by the
anti-magic wavelength $\lambda_{\rm am}/2 \simeq \lambda_0$, so in
this case there will be an additional state independent
potential\cite{gerbier}. We have checked that this does not affect the
appearance of the $\mathbb{Z}_2$ topological insulator phase for the
lowest 4 bands.

While the robust and simple realization of a 3D topological insulator is a key result, the choice \eqref{eq:cossumpure} might seem fortuitous. The remaining part of the Letter introduces a constructive approach for finding 3D systems with  $\mathbb{Z}_2$ nontrivial parities. A hint for this is given by the 2D case, if one notes that the $\bGam_{nm}$ with negative parities are precisely the points in momentum space with fourfold degeneracies.  This can be understood using the following symmetry considerations. 
The Hamiltonian is invariant under translations by the lattice vectors ${\bm a}_i$.  The associated reciprocal lattice vectors are simply $\bm{\kappa}_j$, and define the Brillouin zone for the conserved momentum $\bk$.  Using the Bloch decomposition $\psi_\bk=\exp(i\mathbf{\bk}\cdot\mathbf{\br})
u_\bk(\mathbf{\br})$, the eigenvalue problem at each $\bk$ point follows from
the Hamiltonian
\begin{equation}
\hat{H}_\mathbf{k}= \frac{(\hat{{\bm
      p}}+\hbar{\bm k})^2}{2m} \openone_N + V \hat{M}^\prime({\bm r}) \label{eq:Hk}
\end{equation} 
acting on lattice periodic functions. 
 $\hH_\bk$ has half-translation symmetries $\hT_1=\hSig_2 \hT_{{\bm a}_1/2}$ and
$\hT_2=\hSig_1\hsig_3 \hT_{{\bm a}_2/2}$. (Here $\hT_\mathbf{v}$ is the
operator of translation with the vector $\mathbf{v}$.)  
At the special momenta $\bGam_{nm}$, $\hat{H}_{\mathbf{\Gamma}_{nm}}$ is invariant under both generalized time-reversal and inversion operators:
$\hth_{nm}=\exp(-i2\bGam_{nm}\cdot \br)\hth$, and
$\hP_{nm}=\exp(-i2\bGam_{nm}\cdot \br)\hP$. 
The two symmetries
commute,  hence $u_{\bGam_{nm}}$ and its Kramers partner $\hth_{nm}u_{\bGam_{nm}}$ share the same inversion eigenvalue $\xi_{nm}$.
From the fact that $\hT_i^2=1$ on lattice periodic functions, it follows that \mbox{$\hT_i   \hP_{nm}=\exp(-i\bGam_{nm}\cdot \ba_i)\hP_{nm}\hT_i$}. This means that at $\bGam_{10}$ the application of $\hT_1$, at $\bGam_{01}$ of $\hT_2$ and at $\bGam_{11}$ of either $\hT_i$ reverses the inversion eigenvalue: this
requires, firstly, that the energies are fourfold degenerate, and, secondly, that for each energy the inversion eigenvalues come in opposite pairs\cite{Gam00note}. This explains why the degenerate points contributed with $(-1)$ in Eq.~\eqref{eq:FuKaneinv}. 

These considerations allow one to construct 3D configurations with $\mathbb{Z}_2$ nontrivial parities: consider a model which has a reflection symmetry $\ba_1\leftrightarrow\ba_2$ and a half-translation symmetry by
$\bd=\frac{1}{2}(\ba_1+\ba_2+\ba_3)$. ($\ba_3$ is the third primitive lattice vector.) The reflection symmetry results in
$\xi_{nm1}=\xi_{mn1}$, while the half translation symmetry means that at
$\bGam_{001}$ and $\bGam_{111}$ the inversion eigenvalues come in opposite
pairs for each energy, making the energies fourfold degenerate. Thus, for any $2N$ doublets  we have 
$\prod_{\alpha=1}^{2N}\prod_{nm}\xi^{(\alpha)}_{nm1}=1$, so the net parity is $\prod_{\alpha=1}^{2N}\prod_{nm}\xi^{(\alpha)}_{nm0}$. Assume that the model can be written as
$\hH_\bk=\hH_\bk^{(2D)}+\frac{(\hat{{\bm p}}+\hbar{\bm k})^2_z}{2m} +
\hH_\bk^{(\text{pert})}$. If $\hH_\bk^{(2D)}$ realizes a two dimensional
topological insulator with $2N$ filled Kramers pair of bands, then
$\prod_{\alpha=1}^{2N}\prod_{nm}\xi^{(\alpha)}_{nm0} = -1$  for a weak
$\hH_\bk^{(\text{pert})}$. This means that the net parity
for the lowest 
$2N$ Kramers pairs must be negative:  the model is {\it guaranteed}
to have the parities of a 3D $\mathbb{Z}_2$ topological insulator.
 The choice Eq.~\eqref{eq:cossumpure} is one of the simplest ones consistent with this strategy.

In summary, we have described how simple forms of optical coupling 
can lead to low-energy bands that have non-trivial $\mathbb{Z}_2$
topological invariant, in both 2D and 3D.
Weakly interacting fermions filling these bands will form incompressible band
insulators, and exhibit the expected features of these topological phases:
notably the metallic surface states.
The properties of these surface states
could be probed by studying the (spin-resolved) collective modes of the atomic
gas. Since we work close to the nearly free electron limit, the required temperature scale is set by the recoil energy, which is of the order of \mbox{$0.1$mK} in the case of $^{171}$Yb,  easily accessible in experiments.

The realization of these topological bandstructures in ultracold
atomic gases will allow the study of interesting properties that
cannot readily be explored in solid state systems.  An important
issue concerns the effects of (attractive or repulsive) interactions \cite{clfootnote}.
Weak interactions can affect the properties of the surface modes,
while strong interactions may drive the system into
strongly-correlated topological phases, either with TRS
preserved\cite{zhangint}, or with broken TRS in which case they are
related to fractional quantum Hall states. Furthermore, cold atom
implementations will allow studies of the effects of tuning the nature
of the bands, inducing transitions in their topology.  Introducing a
phase difference between the electric fields in (\ref{eq:coupling})
breaks TRS; this could be used to study quantum phase transitions out
of the $\mathbb{Z}_2$ topological insulator phase, into (for example)
integer quantum Hall states.

\acknowledgments{This work was supported by EPSRC Grant EP/F032773/1.}

\section{Supplementary Material}
In this Supplementary Material, we briefly summarize some basic results from Ref.~\onlinecite{FuKaneinv} about evaluating the $\mathbb{Z}_2$ invariant in inversion symmetric systems.  We repeat them here to help  integrate the key formula, Eq.~\eqref{eq:FuKaneinvsupp} below, into the literature on cold atomic gases. A more elaborate discussion, including detailed derivations can be found in Refs.~\onlinecite{FuKaneinv,hasankane}.

 A time-reversal invariant band-insulator is characterized by a
$\mathbb{Z}_2$ invariant $\nu$, with $\nu=0$ and $\nu=1$ denoting
trivial and non-trivial insulator respectively. The value of this
invariant follows from the properties of the Bloch wavefunctions of
the occupied bands.  For the generic case (with no symmetries other than time-reversal), there are several equivalent ways for its
evaluation. However, all methods require a careful gauge choice for
the wavefunctions across the Brillouin zone. Finding such a gauge
is often difficult.

Shortcuts are possible in the presence of additional symmetries. For example, for 2D systems where the $z$-component of the spin is conserved, evaluating $\nu$ reduces to the calculation of the Chern number $C$ from the wavefunctions of one of the spin components. The result is simply 
\begin{equation}
\nu=C\  \text{mod}\  2.
\end{equation}
This result is used to establish the quantum spin Hall phase of Eq.~\eqref{eq:umu} of the main text for $\delta=0$. 

In the main text, we extensively use another shortcut, which is available if the system has inversion symmetry. This  leads to a radical simplification  in both 2D and 3D. Before citing the result of Ref.~\onlinecite{FuKaneinv}, we briefly summarize some basic facts about the eigenstates of inversion symmetric, time-reversal invariant, lattice periodic Hamiltonians. Inversion symmetry means that the Hamiltonian $\hat{H}(\br)$ is invariant under $\hP:\br\rightarrow -\br$. The Bloch Hamiltonian is 
\begin{equation}
\hat{H}_\mathbf{k}(\br)= e^{-i\bk\br}\hat{H}(\br) e^{i\bk\br},\label{eq:Hksupp}
\end{equation} 
and it acts on lattice periodic functions (Bloch functions). We denote the reciprocal lattice vectors by $\bkappa_j$. At a generic $\bk$, unlike $\hat{H}(\br)$, the Bloch Hamiltonian \eqref{eq:Hksupp} is not invariant under $\hat P$ nor under time-reversal $\hat{\theta}$, since both map $\bk\rightarrow -\bk$. The momenta $\bGam_{\{n_j\}}=\sum_jn_j\bkappa_j/2$ ($n_j=0,1$), however, are special because $\bGam_{\{n_j\}}$ and $-\bGam_{\{n_j\}}$ are equivalent  wavevectors. This means that  $\hat{H}(\br)_{\mathbf{\Gamma}_{\{n_j\}}}$ is
invariant under both generalized time-reversal and inversion operators:
\mbox{$\hth_{\{n_j\}}=\exp(-i2\bGam_{\{n_j\}} \br)\hth$}, and
$\hP_{\{n_j\}}=\exp(-i2\bGam_{\{n_j\}} \br)\hP$, where the exponentials take $H(\br)_{-\bGam_{\{n_j\}}}$ back to $H(\br)_{\bGam_{\{n_j\}}}$  [cf. Eq.~\eqref{eq:Hksupp}]. Note that the operators $\hth_{\{n_j\}}$ and $\hP_{\{n_j\}}$ map lattice periodic functions to lattice periodic functions.

Because of the generalized inversion symmetry, the eigenstates $u(\br)_{\bGam_{\{n_j\}}}$ of $H(\br)_{\bGam_{\{n_j\}}}$ can be labeled by inversion eigenvalues $\xi_{\{n_j\}}=\pm 1$, 
\begin{equation}
\hP_{\{n_j\}}u(\br)_{\bGam_{\{n_j\}}}=\xi_{\{n_j\}}u(\br)_{\bGam_{\{n_j\}}}.
\end{equation}
It is straightforward to show that the corresponding eigenstates $\psi(\br)_{\bGam_{\{n_j\}}}=\exp(i\bGam_{\{n_j\}}\br)u(\br)_{\bGam_{\{n_j\}}}$ of the Hamiltonian $\hat{H}(\br)$  are eigenstates of $\hP$ (and not $\hP_{\{n_j\}}$) with the same eigenvalue $\xi_{\{n_j\}}$.

The symmetry under $\hth_{\{n_j\}}$ means that the  energies at $\bGam_{\{n_j\}}$ (i.e., the eigenvalues of $\hat{H}(\br)_{\bGam_{\{n_j\}}}$) come in Kramers degenerate pairs. The Kramers pair of a Bloch  eigenstate $u(\br)_{\bGam_{\{n_j\}}}$ is $\hth_{\{n_j\}}u(\br)_{\bGam_{\{n_j\}}}$. 
Because the symmetries $\hth_{\{n_j\}}$ and  $\hP_{\{n_j\}}$ commute, $u(\br)_{\bGam_{\{n_j\}}}$ and $\hth_{\{n_j\}}u(\br)_{\bGam_{\{n_j\}}}$   share the same inversion eigenvalue. The $\alpha$-th Kramers doublet at $\bGam_{\{n_j\}}$  can thus be characterized by a single inversion eigenvalue $\xi^{(\alpha)}_{\{n_j\}}$. 

We are now ready to state the result of Ref.~\onlinecite{FuKaneinv}:   Fu and Kane showed that the inversion eigenvalues
corresponding to the filled bands straightforwardly determine whether an
insulator is in a topological phase. The system is a topological insulator  ($\nu=1$)
if \begin{equation} 
\prod_{\{n_j\}}\prod_{\alpha\in\text{filled}}
  \xi^{(\alpha)}_{\{n_j\}}=-1,
\label{eq:FuKaneinvsupp}
\end{equation}
where $\alpha$ runs over the Kramers doublets of the filled bands. Note that the invariant \eqref{eq:FuKaneinvsupp} is even simpler  than the Chern number, since it does not require  the integration of  a (possibly complicated) function over the Brillouin zone;  its sole inputs are  the  parity eigenvalues at  a  few discrete points in momentum space.


\vskip-1cm


\begin{thebibliography}{10}

\bibitem{hasankane}
M.~Z. Hasan and C.~L. Kane, Rev. Mod. Phys. {\bf 82},  3045  (2010).
\bibitem{qizhang}
X.-L. {Qi} and S.-C. {Zhang}, Rev. Mod. Phys. {\bf 83},  1057  (2011).

\bibitem{thoulesschern}
D.~J. Thouless, M. Kohmoto, M.~P. Nightingale, and M. den Nijs, Phys. Rev.
  Lett. {\bf 49},  405  (1982).

\bibitem{KaneMele05FuKaneMele07}
C.~L. Kane and E.~J. Mele, Phys. Rev. Lett. {\bf 95},  146802  (2005);
L. Fu, C.~L. Kane, and E.~J. Mele, Phys. Rev. Lett. {\bf 98},  106803  (2007).

\bibitem{MooreBalents}
J.~E. Moore and L. Balents, Phys. Rev. B {\bf 75},  121306  (2007).

\bibitem{Roy2dRoy3d}
R. Roy, Phys. Rev. B {\bf 79},  195321  (2009);
R. Roy, Phys. Rev. B {\bf 79},  195322  (2009).

\bibitem{blochdz}
I. Bloch, J. Dalibard, and W. Zwerger, Rev. Mod. Phys. {\bf 80},  885  (2008).

\bibitem{lewensteinrev}
M. Lewenstein {\it et~al.}, Advances in Physics {\bf 56},  243  (2007).

\bibitem{ofl}
N.~R. Cooper, Phys. Rev. Lett. {\bf 106},  175301  (2011).

\bibitem{dalibardreview}
J. {Dalibard}, F. {Gerbier}, G. {Juzeli{\= u}nas}, and P. {{\"O}hberg}, arXiv:1008.5378.

\bibitem{dudarevstanescuspielmanso}
A.~M. Dudarev, R.~B. Diener, I. Carusotto, and Q. Niu, Phys. Rev. Lett. {\bf
  92},  153005  (2004);
T.~D. Stanescu, C. Zhang, and V. Galitski, Phys. Rev. Lett. {\bf 99},  110403
  (2007);
Y.-J. Lin, K. Jimenez-Garcia, and I.~B. Spielman,  Nature {\bf 471},  83  (2011).

\bibitem{goldmantopoins}
N. Goldman {\it et~al.}, Phys. Rev. Lett. {\bf 105},  255302  (2010);
%
A. Bermudez {\it et~al.}, Phys. Rev. Lett. {\bf 105},  190404  (2010).

\bibitem{gerbier}
F. Gerbier and J. Dalibard, New J. Phys. {\bf 12},  033007  (2010).

\bibitem{taie}
S. Taie {\it et~al.}, Phys. Rev. Lett. {\bf 105},  190401  (2010).

\bibitem{cohen}
C. Cohen-Tannoudji, J. Dupont-Roc, and G. Grynberg, {\em Atom-Photon
  Interactions} (Wiley, New York, 1992).

\bibitem{llqm}
L.~D. Landau and E.~M. Lifshitz, {\em Quantum Mechanics} (Pergamon Press,
  London, Paris, 1958).

\bibitem{Utrnote}
The time-reversal operator $\hat{\theta}$ is invariant under $\hat{U}$.



\bibitem{kplusfootnote}
Note that ${\bm \kappa}_2$ of Ref.\protect\cite{ofl} is ${\bm \kappa}_1+{\bm
  \kappa}_2$ in the notation here.

\bibitem{FuKaneinv}
L. Fu and C.~L. Kane, Phys. Rev. B {\bf 76},  045302  (2007).

\bibitem{supplmat} See Supplementary Material for additional background information on the Fu-Kane formula for the $\mathbb{Z}_2$ invariant. 

\bibitem{Gam00note}
At $\bGam_{00}$ the fourfold degeneracy is split into two doublets; these two
  doublets can have the same or different inversion eigenvalues.

\bibitem{clfootnote}
To study interaction effects it may help to use an implementation based on hyperfine levels, for which collisional losses can be reduced.

\bibitem{zhangint}
Z. Wang, X.-L. Qi, and S.-C. Zhang, Phys. Rev. Lett. {\bf 105},  256803
  (2010).

\end{thebibliography}
\end{document}